%% file: jfm.tex
\documentclass[]{jfm}

\usepackage{graphicx}
\usepackage{newtxtext}
\usepackage{newtxmath}
\usepackage{natbib}
\usepackage{hyperref}
\usepackage{physics}
\usepackage{caption}
\hypersetup{
    colorlinks = true,
    urlcolor   = blue,
    citecolor  = black,
}

\newcommand{\RomanNumeralCaps}[1]
\linenumbers

\title{Linear instability of viscous parallel shear flows: Revisiting the perturbation no-slip condition}

\author{John O. Dabiri\aff{1,2}
  \corresp{\email{jodabiri@caltech.edu}}
  \and Anthony Leonard\aff{1}}

\affiliation{\aff{1}Graduate Aerospace Laboratories, California Institute of Technology, Pasadena, USA
\aff{2}Mechanical and Civil Engineering, California Institute of Technology, Pasadena, USA}

\begin{document}
\maketitle

\begin{abstract}
Linear stability analysis currently fails to predict turbulence transition in canonical viscous flows. We show that two alternative models of the boundary condition for incipient perturbations at solid walls produce linear instabilities that could be sufficient to explain turbulence transition. In many cases, the near-wall behavior of the discovered instabilities is empirically indistinguishable from the classical no-slip condition. The ability of these alternative boundary conditions to predict linear instabilities that are consistent with turbulence transition suggests that the no-slip condition may be an overly simplified model of fluid-solid interface physics, particularly as a description of the flow perturbations that lead to turbulence transition in wall-bounded flows. 
\end{abstract}


{\bf MSC Codes }  {\it(Optional)} Please enter your MSC Codes here

\section{Introduction}
\label{sec:headings}

A predictive model of the transition from laminar flow to turbulence in viscous parallel shear flows has remained elusive since seminal empirical observations of turbulence transition in the late 19th century~\citep{Reynolds1883}. One of the earliest and most thoroughly explored approaches to this problem examines the linear stability of the Navier-Stokes equations when subjected to small velocity perturbations~\citep{Drazin2004}. Evaluation of the temporal growth of the perturbations typically proceeds by solution of the Orr-Sommerfeld eigenvalue equation (see Appendix A), which depends on the Reynolds number of the flow, $Re = U_0L_0/\nu_0$, where $U_0$ and $L_0$ are characteristic flow speed and length scales, respectively, and $\nu_0$ is the kinematic viscosity of the fluid. Velocity perturbations are constrained to satisfy a no-slip condition at the fluid-solid interfaces. The eigenvalue problem conventionally incorporates this requirement with the homogeneous boundary condition:
    
\begin{equation}\label{eq:noslip}
\mathbf{\tilde{u}}(\mathbf{x_{wall}},t) = 0
\end{equation}
    
\noindent where $\mathbf{\tilde{u}}$ is the perturbation velocity vector amplitude and $\mathbf{x_{wall}}$ is the location of the fluid-solid interfaces.

The aforementioned analysis predicts that plane Couette flow is stable to all two-dimensional linear perturbations at all Reynolds numbers~\citep{Davey1973}. This prediction is contradicted by empirical observations of transition to turbulence at channel Reynolds numbers as low as $Re \approx 360$~\citep{Drazin2004}. The analysis method similarly finds no linear instability to account for the observed transition of Hagen-Poiseuille pipe flow to turbulence at a Reynolds number $Re \approx 2000$. For planar Poiseuille flow, linear stability analysis does identify a single unstable eigenmode, which first appears at a Reynolds number $Re \approx 5772$~\citep{Orszag1971}. In practice, however, transition to turbulence is observed to occur at significantly lower Reynolds numbers $Re \approx 1000$. Moreover, the growth rate associated with the unstable eigenvalue $\hat{\omega}$ at $Re \approx 5772$ is relatively weak, e.g., its imaginary part $\mathbb{I}[\hat{\omega}] \approx 0.0037$ for $\alpha \approx 1.02$. The associated unstable eigenmode may therefore be insufficient to trigger turbulence.

This apparent inability of linear stability theory to accurately predict turbulence transition in a variety of viscous parallel shear flows has motivated the exploration of alternative frameworks to explain and possibly predict turbulence transition. These include consideration of finite-sized velocity perturbations~\citep{Orszag1980}, nonlinear transition processes~\citep{Schmid2001}, and transient growth mechanisms such as those associated with non-normality of the Orr-Sommerfeld eigenvalue equation~\citep{Trefethen1993}. While it is beyond the scope of this paper to comprehensively review the rich literature on the theory of turbulence transition, the reader is referred to an excellent recent review article~\citep{Avila2023}.

The aforementioned body of work suggests that linear instability may not be a necessary condition for turbulence transition. Nonetheless, in this paper we show that linear stability analysis is \emph{sufficient} to quantitatively predict the occurrence of flow instabilities in each of three canonical viscous parallel shear flows: plane Couette flow, plane Poiseuille flow, and Hagen-Poiseuille pipe flow. We proceed by considering two alternative models for incipient flow perturbations in these canonical flows. In the first case, we use a more general ansatz for the perturbation boundary condition in the Orr-Sommerfeld equation, of which the no-slip condition is a limiting case. In the second case, we model the perturbation behavior at the wall using a variation of Stokes' second problem~\citep{Landau1987}. While the latter model represents a more significant departure from the conventional perturbation no-slip condition, it leads to physically realistic, quantitative predictions of a Reynolds number dependent transition to linear instability that is consistent with empirical observations, i.e., $Re_{crit} \sim \mathcal{O}(10^2-10^4)$.

Both of the alternative models for incipient perturbations exhibit near-wall behavior that is, in many cases, empirically indistinguishable from the classical no-slip condition. The superior predictive capability of these new models---without the need to appeal to nonlinear or transient growth processes---suggests a re-examination of the no-slip condition as a sufficiently accurate model of the fluid-solid interface physics. More generally, these results hint at the possibility that more detailed study of flow physics at fluid-solid interfaces could lead to a better understanding of turbulence transition.

Section 2 explores the first alternative model for incipient perturbations. This model includes the conventional no-slip condition as a limiting case, enabling more direct comparison with prior results in the literature. While this feature is pedagogically useful, this perturbation model exhibits an associated trade-off, namely, that the predicted linear instability does not exhibit an explicit Reynolds number dependence. Section 3 examines the second alternative model, which represents a more significant departure from the form of no-slip condition conventionally applied to velocity perturbations in linear stability analysis. This model does successfully predict a Reynolds number dependent linear instability that is consistent with empirical observations of turbulence. Section 4 concludes the paper by discussing the implications of these findings for our understanding of the role of fluid-solid interface physics in prediction of turbulence transition.

\section{Perturbation Model I}
\subsection{Generalized boundary condition}
\label{sec:headings}

Let us replace the streamwise component of the no-slip condition in equation (\ref{eq:noslip}) with a more general, homogeneous boundary condition that depends on the shear rate of the velocity perturbation at the wall:

   \begin{equation} \label{eq:Redependentslip}
    \tilde{u}(\mathbf{x_{wall}},t) \mp S\, \tilde{u}'(\mathbf{x_{wall}},t) = 0
    \end{equation}

\noindent where $\tilde{u}$ is the streamwise component of the velocity perturbation amplitude, $\tilde{u}' = \partial \tilde{u}/\partial y$ for the plane Couette and Poiseuille flows, $\tilde{u}' = \partial \tilde{u}/\partial r$ for the Hagen-Poiseuille pipe flow, $S$ is a characterisic slip length constant (normalized by the channel half-width or pipe radius), and the sign $\mp$ applies to the wall at $y$ (or $r$) $= 1$ and $y = -1$, respectively.

Physically, the boundary condition in equation (\ref{eq:Redependentslip}) requires that any non-zero perturbation velocity at the wall is in the same direction as the corresponding shear exerted on the fluid by the wall. While it is similar in form to the Navier slip boundary condition~\citep{Lauga2007}, there are two key differences between this model and prior studies of the effect of wall slip on linear stability (e.g.,~\cite{Lauga2005, Chai2019, Ceccacci2022}). First, in the present analysis the boundary condition (\ref{eq:Redependentslip}) applies only to velocity \emph{perturbations}; the base flow is assumed to exhibit the conventional no-slip condition. Physically, this assumption is consistent with the ansatz that the distribution of velocity perturbations exhibits a zero mean, and therefore the velocity perturbations do not change the base flow. By contrast, in the aforementioned studies the base flow is assumed to exhibit wall slip even in the absence of velocity perturbations. While the assumption of non-zero base flow slip is appropriate in the context of those prior studies, e.g., where the walls were assumed to exhibit hydrophobicity or other surface treatment, the goal of the present study is to examine wall-bounded flows more generally.

The second key distinction between the present perturbation model and prior studies is that the perturbation velocity at the wall is applied here in the same direction as the corresponding shear exerted on the fluid by the wall. In contrast, the previous studies apply perturbation wall slip in the \emph{opposite} direction of the wall shear exerted on the fluid. Physically, the present model of wall slip assumes that the velocity perturbations are caused by wall shear perturbations on the adjacent fluid. This distinction is significant, and we confirmed that the stability predictions that follow do indeed depend on relative directions of the velocity perturbation and the wall shear on the fluid (see Appendix E).  

This generalized boundary condition (\ref{eq:Redependentslip}) reduces to the no-slip condition (\ref{eq:noslip}) when the slip parameter $S = 0$. The no-slip condition is also satisfied by velocity perturbations with non-zero values of the slip parameter $S$ if the corresponding wall-normal gradient of streamwise flow speed is zero at the wall, i.e., $\tilde{u}'(\mathbf{x_{wall}},t) = 0$. While we presently focus on the case of real-valued slip parameter $S$, we have observed similar results for complex $S$. In the latter case, a phase difference exists between the perturbation wall slip and the perturbation wall shear.

\subsection{Linear stability maps}
\label{sec:headings}

The Orr-Sommerfeld eigenvalue equation was solved for plane Couette flow, plane Poiseuille flow, and Hagen-Poiseuille pipe flow using Chebyshev collocation \citep{Schmid2001, Malik2019}; see Appendix A for governing equations and data repository for Matlab implementations. The boundary condition in equation (\ref{eq:Redependentslip}) was evaluated for values of the slip parameter $S$ = 0 and ranging from $S = 10^{-7}$ to $S = 1$ in 10 equally-spaced increments per decade. The Reynolds number was varied from $Re = 50$ to $Re = 10,000$ in increments of 10.

\begin{figure}
    \centering
    \includegraphics[width=\textwidth]{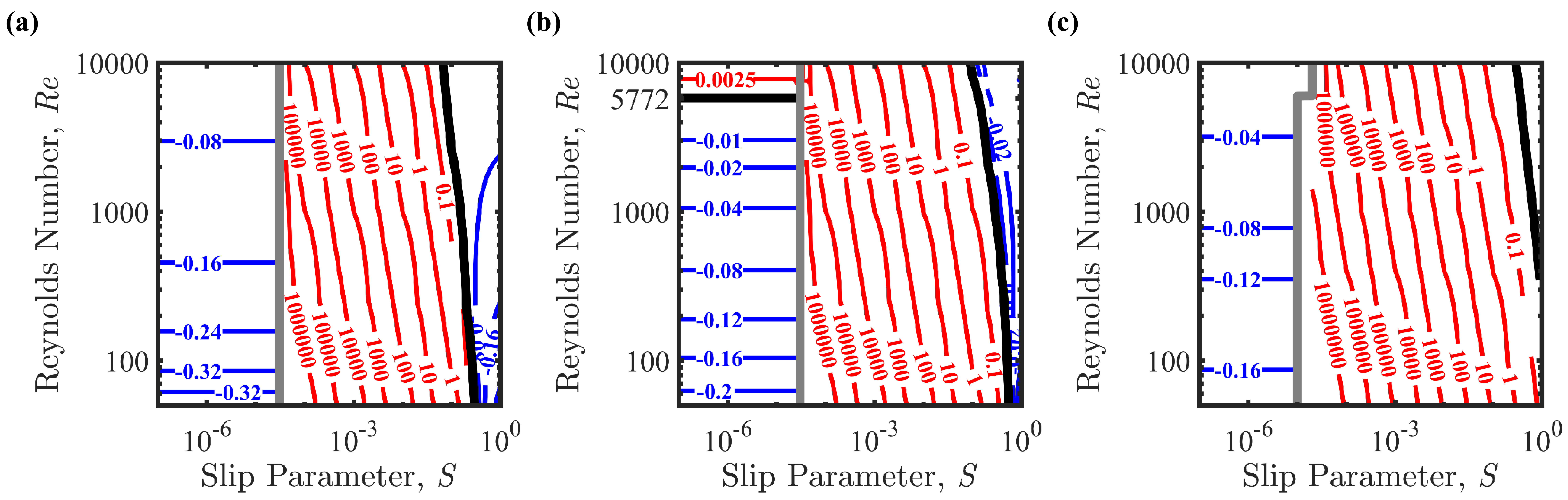}
    \caption{Contour maps of the maximum Orr-Sommerfeld eigenvalue imaginary part $\mathbb{I}[\hat{\omega}]$ versus slip parameter $S$ and Reynolds number $Re$ for (a) plane Couette flow, (b) plane Poiseuille flow, and (c) Hagen-Poiseuille pipe flow. Blue contours indicate regions of linear stability, and red contours indicate regions of linear instability. Black contours indicate neutral stability boundaries. The Reynolds number $Re \approx 5772$ of the single unstable eigenmode in plane Poiseuille flow for $S = 0$ is indicated on the ordinate axis in panel (b). Vertical gray lines correspond to a discontinuous change in predicted hydrodynamic stability, reflecting the inability of the Chebyshev expansion ($N = 400$) to resolve unstable eigenmodes for values of slip parameter $S$ below the gray line. Wavenumbers are $(\alpha, \beta) = (1, 0)$ for the planar flows and wavenumbers $(\alpha, n) = (1, 1)$ for Hagen-Poiseuille pipe flow. See Appendices A and B for details of each calculation.}
    \label{fig:maps}
\end{figure}

For combinations of slip parameter $S$ and Reynolds number $Re$ spanning this range, figure \ref{fig:maps} plots contours of the maximum eigenvalue imaginary part, i.e., $\mathbb{I}[\hat{\omega}]$, for Orr-Sommerfeld solutions corresponding to plane Couette flow, plane Poiseuille flow, and Hagen-Poiseuille pipe flow, respectively (Similar contour maps of the second-largest eigenvalue are provided in Appendix C for reference).

\begin{figure*}
   \centerline{\includegraphics[width=0.95\textwidth]{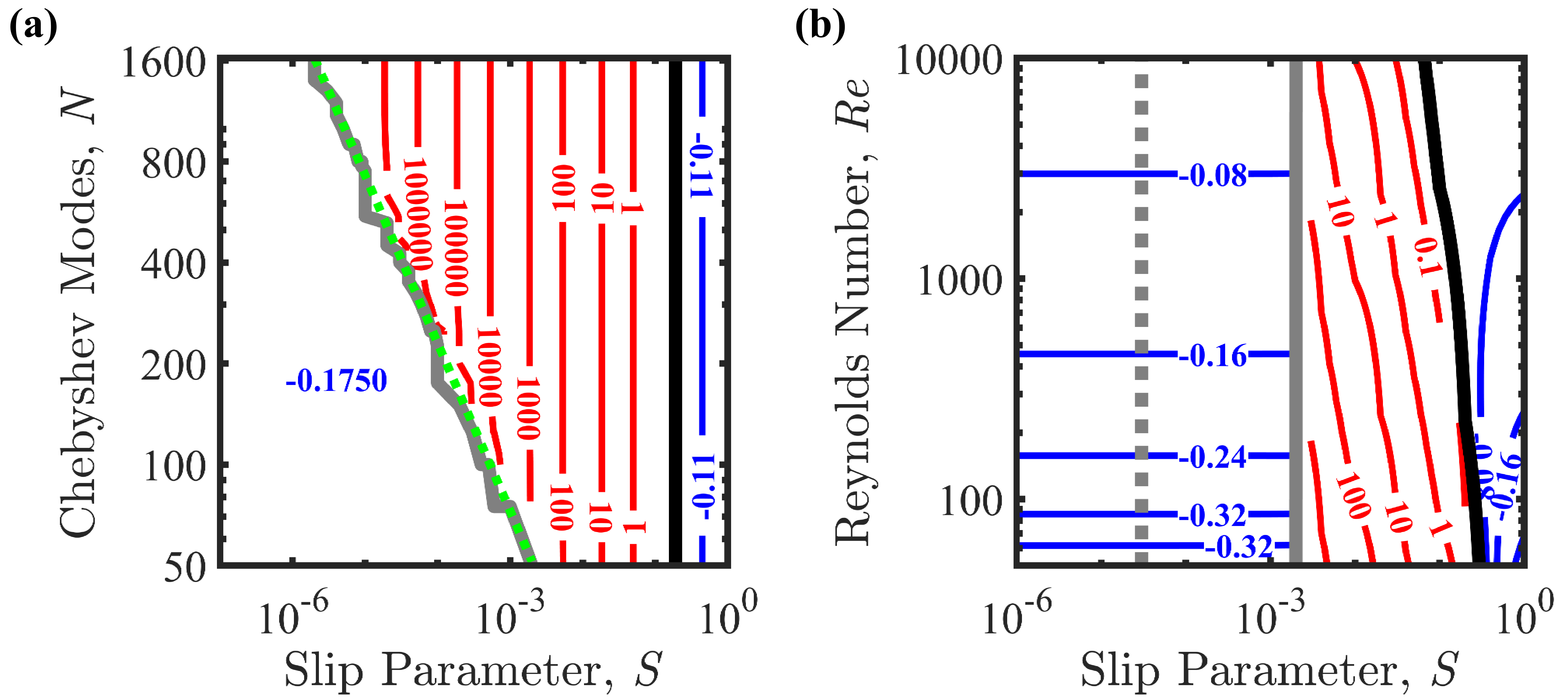}}
    
    \caption{(a) Contour map of the maximum Orr-Sommerfeld eigenvalue imaginary part $\mathbb{I}[\hat{\omega}]$ versus slip parameter $S$ and number of Chebyshev modes $N$ for plane Couette flow at $Re = 360$ and wavenumbers $(\alpha, \beta) = (1, 0)$. Blue contours indicate regions of linear stability, and red contours indicate regions of linear instability. Neutral stability boundary is indicated by black line. Discontinuity in stability is indicated by gray curve. The location of the stability discontinuity shifts to lower values of the slip parameter $S$ as $N$ increases, reflecting the ability to resolve unstable eigenmodes with decreasing slip length scale $S$ for increasing numbers of modes $N$ used in the Chebyshev expansion. The minimum number of Chebyshev modes $N_{min}$ required to resolve unstable eigenmodes for a given slip parameter $S$ is well approximated by $N_{min} \approx 2.3\,S^{-1/2}$ (i.e., green dotted line).  (b) Contour map of the maximum Orr-Sommerfeld eigenvalue imaginary part versus slip parameter $S$ and Reynolds number $Re$ computed using $N = 50$ Chebyshev modes. The stability discontinuity (i.e., gray dotted line) from figure \ref{fig:maps}(a), computed using $N = 400$ Chebyshev modes, is reproduced to illustrate the rightward shift in the threshold value of the slip parameter $S_{crit}$ corresponding to the stability discontinuity when fewer Chebyshev modes are used.}  
    \label{fig:couette_Ndependence_combined}
\end{figure*}

For values of slip parameter $S$ approaching zero, the linear stability maps are each consistent with the results of conventional linear stability analyses using equation (\ref{eq:noslip}) as the boundary condition for velocity perturbations. Specifically, both the plane Couette flow and the Hagen-Poiseuille pipe flow indicate linear stability of the flow (i.e., $\mathbb{I}[\hat{\omega}] < 0$)  for all Reynolds numbers investigated. The plane Poiseuille flow is also in agreement with previous studies for $S$ approaching zero, with a single unstable eigenmode appearing between $Re = 5770$ and $Re = 5780$~\citep{Orszag1971}.

However, a striking discontinuity is observed in each of the linear stability maps as the slip parameter $S$ is increased above a critical threshold (i.e., vertical gray lines). For values of slip parameter greater than this threshold $S_{crit}$, we observe a region of linear instability spanning up to four decades in $S$. These strongly unstable eigenmodes---with growth rates $\mathbb{I}[\hat{\omega}]$ exceeding $\mathcal{O}(10^7)$ in some cases---appear at all Reynolds numbers investigated, making them potentially relevant to the process of turbulence transition.

The apparent lack of unstable eigenmodes for values of slip parameter $S < S_{crit}$ is an artifact of the limited spatial resolution of the Chebyshev expansion using a finite number of modes. To illustrate this, figure \ref{fig:couette_Ndependence_combined}(a) plots contours of the maximum Orr-Sommerfeld eigenvalue imaginary part $\mathbb{I}[\hat{\omega}]$ versus slip parameter $S$ and number of Chebyshev modes $N$, for the case of plane Couette flow at $Re = 360$ and wavenumbers $(\alpha, \beta) = (1, 0)$. Eigenvalue contours to the right of the gray curve are oriented vertically, indicating that the corresponding eigenvalues are insensitive to the number of Chebyshev modes used to solve the Orr-Sommerfeld equation. This suggests that the computed eigenmodes are not spurious artifacts of the numerical method (cf.~\citet{Dawkins1998}). By contrast, the location of the gray stability discontinuity shifts to lower values of the slip parameter $S$ as the number of Chebyshev modes used to compute the eigenmodes increases. This boundary reflects the finite spatial resolution of the Chebyshev representation of unstable eigenmodes for finite mode numbers $N$. As the number of Chebyshev modes is increased, unstable eigenmodes with smaller slip length scale $S$ are successfully resolved. The minimum number of Chebyshev modes $N_{min}$ required to resolve unstable eigenmodes for a given slip parameter $S$ is well approximated by $N_{min} \approx 2.3\,S^{-1/2}$ (i.e., green dotted line). 

Figure \ref{fig:couette_Ndependence_combined}(b) provides a complementary illustration of this trend, showing the linear stability map computed using $N = 50$ Chebyshev modes. For comparison with the stability map computed using $N = 400$ Chebyshev modes in figure \ref{fig:maps}(a), the stability discontinuity (i.e., gray dotted line) from that plot is reproduced here. The results highlight the need to implement Chebyshev collocation with a sufficient number of collocation points to resolve the distinct shape of unstable eigenmode profiles in close proximity to the wall. Calculations using an insufficient number of Chebyshev modes erroneously predict the absence of any unstable modes. In light of the trend $N_{min} \sim S^{-1/2}$, the number of Chebyshev modes required to resolve unstable eigenmodes becomes prohibitive as the conventional no-slip condition is approached (i.e., $S \rightarrow 0$). To be sure, a similar resolution challenge will be faced by other numerical methods.

\subsection{Asymptotic analysis of Couette flow}
\label{sec:headings}

To further analyze the discovered unstable eigenmodes while circumventing the aforementioned numerical resolution challenges, we conducted an asymptotic analysis of plane Couette flow in the limit of strongly unstable eigenmodes, i.e., $\mathbb{I}[\hat{\omega}] \gg 1$ (see Appendix D). This analysis predicts that the unstable eigenmode growth rate scales with slip parameter $S$ and Reynolds number $Re$ as $\mathbb{I}[\hat{\omega}] \sim S^{-2}\,Re^{-1}$. Moreover, the unstable plane Couette eigenmode amplitude components have the predicted asymptotic form

\begin{equation}\label{eq:unstablemode_u}
\begin{split}
& \tilde{u}(y) \approx (i/\alpha)\Big[(1/S)e^{(y-1)/S} + (1/S)e^{-(y+1)/S} \\
& - \frac{\alpha\cosh \alpha(y+1)}{\sinh 2\alpha}
- \frac{\alpha\cosh \alpha(y-1)}{\sinh 2\alpha}\Big] + \mathcal{O}(S) \\
\end{split}
\end{equation}

\newcommand{\mycomment}[1]{}
\mycomment{}
\begin{equation}\label{eq:unstablemode_v}
\begin{split}
& \tilde{v}(y) \approx e^{(y-1)/S} - e^{-(y+1)/S} - \frac{\sinh \alpha(y+1)}{\sinh 2\alpha} \\
& - \frac{\sinh \alpha(y-1)}{\sinh 2\alpha} + \mathcal{O}(S) \\
\end{split}
\end{equation}

\begin{figure*}
    \centering
    \includegraphics[width=\textwidth]{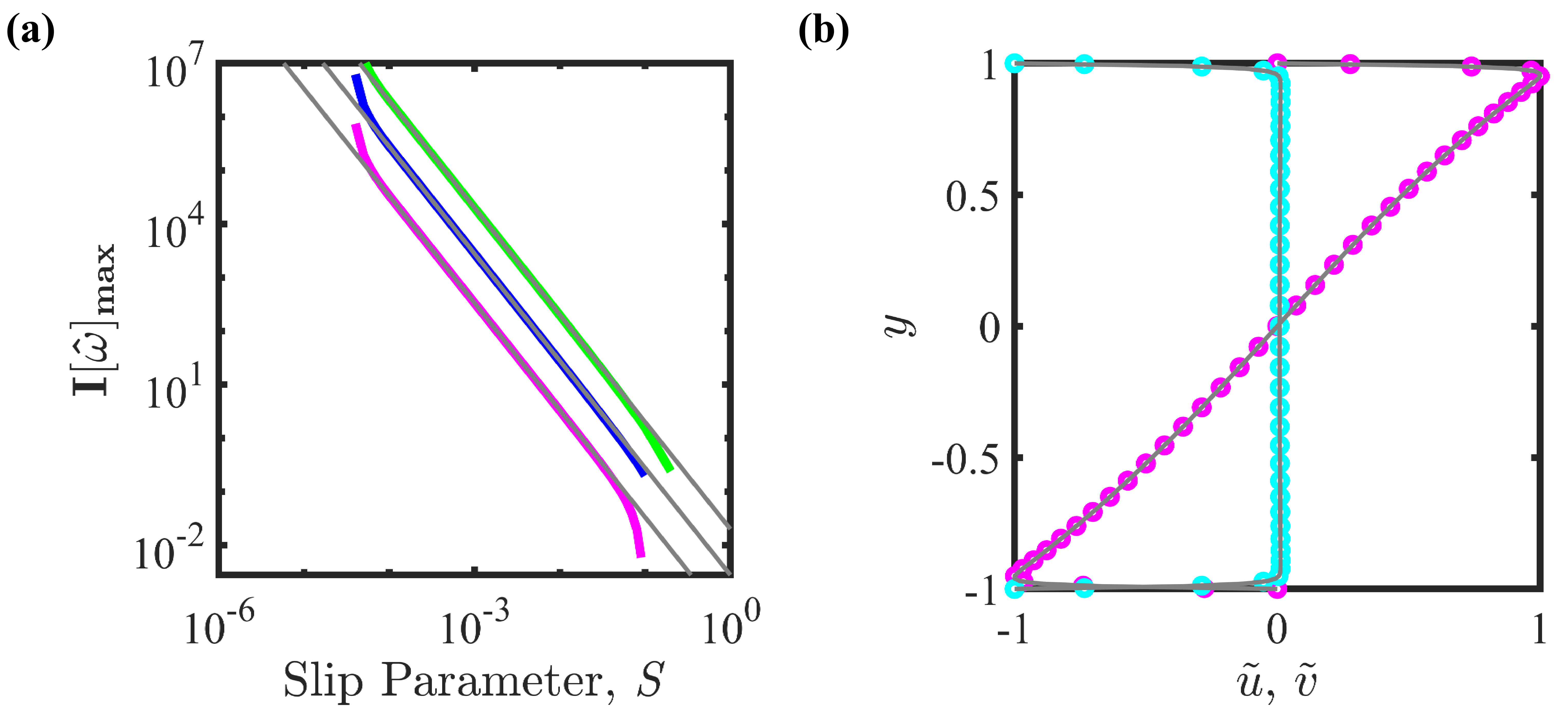}
    \caption{(a) Maximum unstable eigenvalues of plane Couette flow plotted versus slip parameter $S$. Thick colored curves are computed using Chebyshev collocation ($N = 400$) for $Re = 50$ (green), $Re = 360$ (blue), and $Re = 3000$ (magenta). Thin gray lines are corresponding analytical model prediction $\mathbb{I}[\hat{\omega}] \sim S^{-2}\,Re^{-1}$. (b) Streamwise ($\tilde{u}$, cyan) and wall-normal ($\tilde{v}$, magenta) components of unstable eigenmode amplitude profiles computed using Chebyshev collocation ($N = 400$, $Re = 360$, $S = 1 \times 10^{-2}$, and $(\alpha, \beta) = (1, 0)$). Corresponding eigenmode profiles predicted by asymptotic analyses in equations (\ref{eq:unstablemode_u}) and (\ref{eq:unstablemode_v}) are shown in superimposed thin gray curves.}  
    \label{fig:couette_model_combined}
\end{figure*}

Figure \ref{fig:couette_model_combined} compares the predictions of the asymptotic analysis with the calculations of plane Couette flow using Chebyshev collocation with $N = 400$. The agreement is excellent for both the maximum unstable eigenvalues (panel (a)) and the shape of each component of the unstable eigenmode (panel (b)). These results further support the conclusion that the discovered unstable eigenmodes are not spurious numerical artifacts.

Nonetheless, the predicted scaling of the eigenmode growth rate $\mathbb{I}[\hat{\omega}] \sim S^{-2}\,Re^{-1}$ is counterintuitive, as it suggests stronger instability at lower Reynolds numbers. Moreover, the analysis does not exhibit a critical Reynolds number below which instability is not predicted to occur. This prediction is contradicted by the known stability of viscous parallel shear flows at sufficiently low Reynolds number. We speculate that the apparent inconsistency can be resolved by consideration of the predicted shape of the unstable eigenmodes. For example, from equation (\ref{eq:unstablemode_u}) we see that the magnitude of the streamwise component of the unstable eigenmode velocity gradient at the wall goes as $\norm{\partial\tilde{u}/\partial y} \sim \mathbb{I}[\hat{\omega}]Re \gg 1$. In a fluid with finite dynamic viscosity $\mu$, the corresponding wall shear required to create this unstable perturbation is therefore $\tilde{\tau}_{wall} = \mu (\partial\tilde{u}/\partial y) \gg 1$, which may not be physically realizable. Hence, although the analysis predicts that the flow is linearly unstable to eigenmodes of the shape given in equations (\ref{eq:unstablemode_u}) and (\ref{eq:unstablemode_v}), that prediction of eigenmode growth is only relevant if those eigenmodes are actually present in the flow. In practice, the viscosity of real fluids may prevent the appearance of these unstable eigenmodes at low Reynolds numbers due to the large wall shear required to produce them. To be sure, a nonlinear analysis would be necessary to conclusively resolve this inconsistency. 

In addition, not all values of slip parameter $S$ will be physically realizable at all Reynolds numbers. Reynolds-number-dependent slip is commonly observed in flow over complex interfaces, e.g., porous media~\citep{Beavers1967, Wu2018,Guo2020}. If a similar phenomenon is associated with flow \emph{perturbations} over solid walls (i.e., as opposed to the base flow), then the Reynolds number dependence of the present stability predictions may be implicit in the slip parameter $S$.

\section{Perturbation Model II}
\subsection{Further departure from the no-slip condition}

Perturbation model I was formulated with intent of parameterizing modest departures from the no-slip condition, such that the conventional no-slip condition (\ref{eq:noslip}) is a limiting case.  Here we consider the implications of more significant departures from the classical no-slip condition as applied to velocity \emph{perturbations} in linear stability analyses of viscous parallel shear flows. In the analysis that follows, we maintain the assumption that the base flow satisfies the no-slip condition.

Following \cite{Schmid2001}, we consider velocity perturbations of the general Cartesian form

\begin{equation}\label{eq:SHperturbation}
\mathbf{u}(x,y,z,t) = \mathbb{R}[\mathbf{\tilde{u}}(y)e^{i(\alpha x + \beta z - \omega t)}]
\end{equation}

\noindent where the Cartesian components of the perturbation velocity vector are $\mathbf{u} = u \mathbf{\hat{i}} + v \mathbf{\hat{j}} + w \mathbf{\hat{k}}$ in the streamwise (i.e., $\mathbf{\hat{i}}$), wall-normal (i.e., $\mathbf{\hat{j}}$), and transverse (i.e., $\mathbf{\hat{k}}$) directions. 
The temporal evolution of the perturbations is examined by treating the spatial wavenumbers as $\alpha,\beta \in \mathbb{R}$ and the frequency as $\omega = \alpha c$, where $c \in \mathbb{C}$ is the complex phase speed of the perturbation. We will consider the case $\beta = 0$ below without loss of generality.

If we permit non-zero values of the streamwise perturbation velocity magnitude at the wall, the corresponding fluid particle trajectories $X(t)$ along the wall are given by

\begin{equation}\label{eq:trajectoryvel}
\frac{\partial X}{\partial t} = \norm{\mathbf{\tilde{u}_{wall}}}\cos(\alpha x - \omega_R t)e^{\omega_I t}
\end{equation}

\noindent where $\omega_R = \mathbb{R}[\omega]$, $\omega_I = \mathbb{I}[\omega]$, and the phase of $\mathbf{\tilde{u}_{wall}}$ is omitted without loss of generality. Inspection of equation (\ref{eq:trajectoryvel}) shows that fluid particles at the wall are stationary relative to the wall---and therefore satisfy the no-slip condition---where

\begin{equation}\label{eq:stationary1wave}
\alpha x - \omega_R t = \frac{\pi}{2} + \pi j
\end{equation}

\noindent or, equivalently

\begin{equation}\label{eq:stationary1wave}
x = \mathbb{R}[c] t + \frac{(\frac{\pi}{2} + \pi j)}{\alpha}
\end{equation}

\noindent where $j$ is an integer. For single eigenmodes, the discrete locations where the no-slip condition is satisfied are not fixed, but instead propagate with the phase speed $\mathbb{R}[c]$ of the associated eigenmode. Consequently, all fluid particles on the wall will eventually experience non-zero slip. By contrast, for paired eigenmodes with real part $\pm \omega_R$, the velocity of fluid particles at the wall is given by

\begin{equation}\label{eq:trajectoryvel2wave}
\frac{\partial X}{\partial t} = \norm{\mathbf{\tilde{u}_{wall}}}\Big[\cos(\alpha x - \omega_R t) + \cos(\alpha x + \omega_R t)\Big]e^{\omega_I t}
\end{equation}

The term in brackets in equation (\ref{eq:trajectoryvel2wave}) can be rewritten as

\begin{equation}\label{eq:trajectoryvel2wavecosines}
\cos(\alpha x - \omega_R t) + \cos(\alpha x + \omega_R t) = 2\cos(\alpha x)\cos(\omega_R t)
\end{equation}

Hence, the no-slip condition is satisfied at all times at the discrete, fixed locations $x = (\frac{\pi}{2} + \pi j)/\alpha$. Once fluid particles are advected to these locations along the wall, they asymptotically satisfy the no-slip condition thereafter. During any initial transient wall slip to reach these locations, the maximum fluid particle displacement along the wall is $\mathcal{O}(\pi/\alpha)$, i.e., the distance between adjacent locations where $\frac{\partial X}{\partial t} = 0$. The duration of the initial wall slip transient is $\mathcal{O}(\pi/\norm{\mathbf{\tilde{u}_{wall}}}\alpha)$. 

For stable eigenmodes, i.e., $\omega_I < 0$, the wall slip decays exponentially per equation (\ref{eq:trajectoryvel2wave}), rendering any perturbation-induced motion of fluid particles along the wall likely imperceptible in practice. For unstable eigenmodes, the exponentially increasing wall slip speed associated with $\omega_I > 0$ will advect particles to the stationary nodes on the wall more rapidly, thereby hastening their asymptotic approach to the no-slip condition in those cases. To be sure, there will always remain fluid particles in motion along some portion of the wall in order to satisfy continuity. However, the linearity of the eigensolutions implies that this motion can theoretically be arbitrarily small at the incipient stage of an instability. More specifically, if $\tilde{\mathbf{u}}$ is an eigenmode of the Orr-Sommerfeld equations, then any constant multiple $\gamma$ of the eigenmode (i.e., $\gamma\tilde{\mathbf{u}}$), is also an eigenmode, including those for which $\gamma \ll 1$. Together, these considerations lead to the interesting possibility that the wall slip associated with both stable and unstable velocity perturbations will be empirically indistinguishable from the conventional no-slip condition, even if the perturbations do not formally satisfy the no-slip boundary condition.

If the no-slip condition is not the only plausible model for the physics of velocity perturbations in viscous parallel shear flows, then exploration of other physically-realistic boundary conditions for linear stability analysis is warranted. Perturbation Model I was one such alternative. In the following sections, we explore another boundary condition that not only produces linearly unstable eigenmodes, but also leads to quantitative predictions of an explicit critical Reynolds number range associated with linear instability.

\subsection{Womersley model of velocity perturbations}

The Orr-Sommerfeld equation is a fourth-order system, and therefore we must specify four boundary conditions on the two walls of the plane Couette or Poiseuille flow. The first two boundary conditions prohibit flow through the walls:

\begin{equation}\label{eq:nothroughflow1}
\tilde{v}(y = 1,t) = 0
\end{equation}

\begin{equation}\label{eq:nothroughflow2}
\tilde{v}(y = -1,t) = 0
\end{equation}

In lieu of the conventional no-slip condition, we instead hypothesize that the fundamentally oscillatory nature of the streamwise velocity perturbations (cf. equation (\ref{eq:SHperturbation})) can be modeled with a boundary condition based on Stokes' second problem \citep{Landau1987} as

\begin{equation}\label{eq:stokessecond}
\frac{\partial \tilde{u}}{\partial t} - \frac{1}{Re} \frac{\partial^2 \tilde{u}}{\partial y^2}  = 0
\end{equation}

\noindent Equation (\ref{eq:stokessecond}) provides the remaining two required boundary conditions at $y = \pm 1$. Note that equations (\ref{eq:stokessecond}) and (\ref{eq:Redependentslip}) are only two of many physically-motivated ansatzes that could be explored in general. Indeed, a goal of this paper is to demonstrate that there exist boundary conditions in addition to the no-slip condition that may also accurately represent incipient perturbations, and that these alternative boundary conditions may be associated with linear stability that is more consistent with empirical observations of turbulence transition.   

Substituting the form of velocity perturbation in equation (\ref{eq:SHperturbation}) into equation (\ref{eq:stokessecond}), and using $\tilde{u} = (i/\alpha)\mathcal{D}\tilde{v}$ from the continuity equation (see Appendix A), the boundary condition can be expressed as

\begin{equation}\label{eq:stokessecond_tildev}
\omega\frac{\partial \tilde{v}}{\partial y} - \frac{i}{Re}\frac{\partial^3 \tilde{v}}{\partial y^3}  = 0
\end{equation}

We introduce the \emph{perturbation Womersley number}, $\widetilde{Wo}$, as a measure of the transient inertial force associated with the velocity perturbation $\mathbf{u}$ relative to its associated viscous force (cf. \cite{Womersley1955}):

\begin{equation}\label{eq:perturbwomersley}
\widetilde{Wo} = \frac{1}{\alpha}\sqrt{\frac{\omega}{\nu_0}}
\end{equation}

\noindent where $\nu_0$ is the kinematic viscosity of the fluid. The frequency $\omega$ is complex in general; however, we make the simplifying assumption $\norm{\mathbb{I}[\omega]} \ll 1$, from which it follows that  
$\widetilde{Wo} \in \mathbb{R}$. This approximation holds identically for neutral stability, i.e., $\mathbb{I}[\hat{\omega}] = 0$. We will see in the following section that this approximation also captures large regions of parameter space away from the neutral stability boundaries. 

Since the planar flows under present consideration are defined with $U_0 = 1$ and $L_0 = 1$, the boundary condition \ref{eq:stokessecond_tildev} can be written as

\begin{equation}\label{eq:womersleybc}
\frac{\widetilde{Wo}^2\alpha^2}{Re}\frac{\partial \tilde{v}}{\partial y} - \frac{i}{Re}\frac{\partial^3 \tilde{v}}{\partial y^3}  = 0
\end{equation}

 Linear stability analysis proceeds by solving the Orr-Sommerfeld equation using the boundary conditions (\ref{eq:nothroughflow1}), (\ref{eq:nothroughflow2}), and (\ref{eq:womersleybc}) at $y = \pm1$. Here, we use Chebyshev collocation to solve the Orr-Sommerfeld eigenvalue equation following  \cite{Schmid2001} for plane Couette and Poseuille flow, and following \cite{Malik2019} for Hagen-Poiseuille pipe flow (see Appendices A and B).

\subsection{Linear stability maps}
 
\begin{figure}
    \centering
    \includegraphics[width=0.8\textwidth]{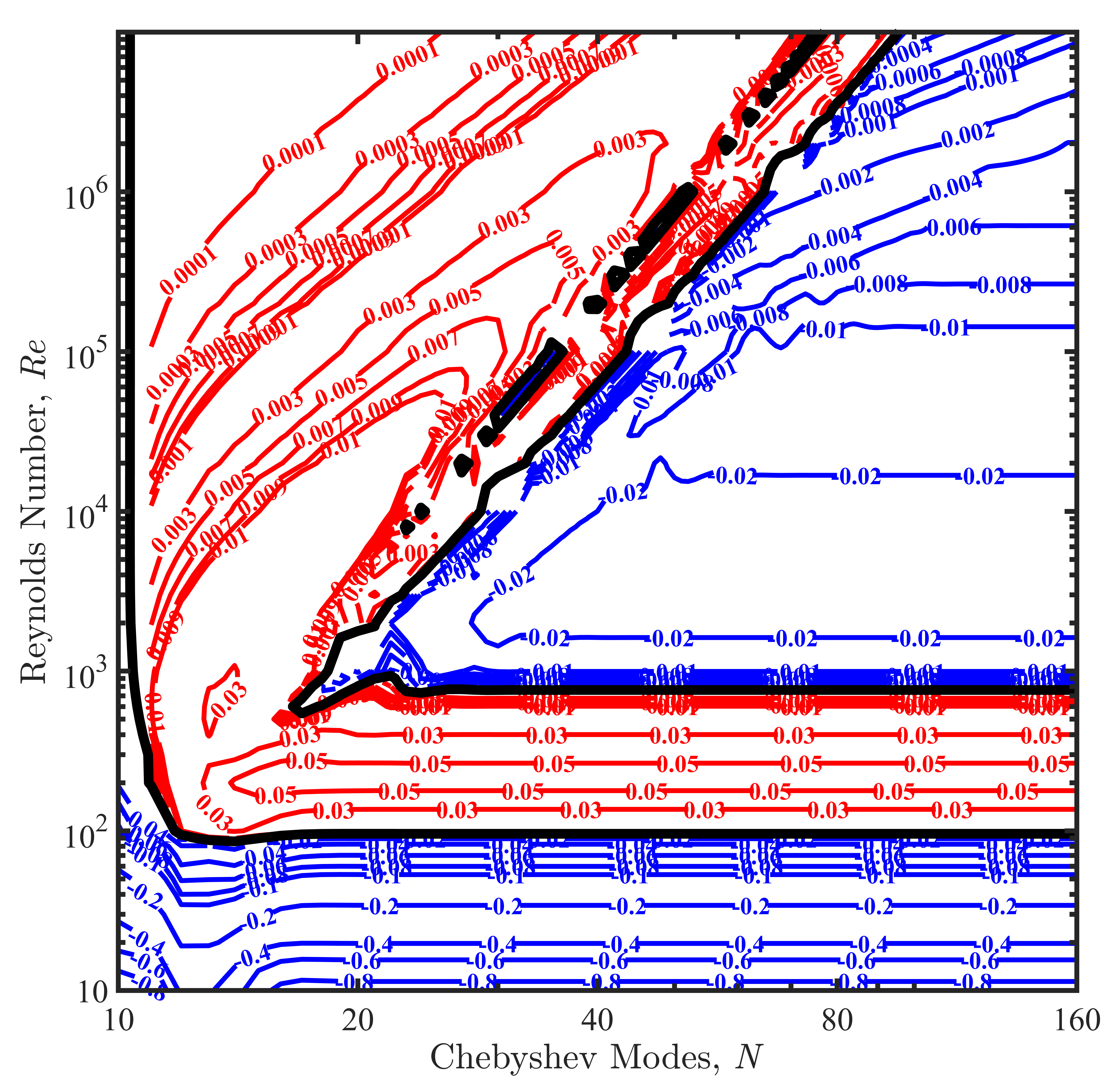}
    \caption{Contour map of the maximum Orr-Sommerfeld eigenvalue imaginary part $\mathbb{I}[\hat{\omega}]$ versus the number of Chebyshev modes $N$ and and Reynolds number $Re$ for plane Couette flow with $\widetilde{Wo} = 10$ and $(\alpha, \beta) = (1, 0)$.} 
    \label{fig:map_Re_N}
\end{figure}

 Figure \ref{fig:map_Re_N} plot contours of the maximum eigenvalue imaginary part, i.e., $\mathbb{I}[\hat{\omega}]$, for Orr-Sommerfeld solutions corresponding to plane Couette flow with $\widetilde{Wo} = 10$ and $(\alpha, \beta) = (1, 0)$. For relatively low numbers of Chebyshev modes $N$ used to solve the Orr-Sommerfeld equation, we observe spurious unstable eigenmodes at increasing Reynolds numbers (i.e., upper left corner of contour plot). The spurious nature of these eigenvalues is evidenced by the dependence of their value on the number of Chebyshev modes used to solve the Orr-Sommerfeld equation \citep{Dawkins1998}. However, as the number of Chebyshev modes is increased, spurious eigenmodes disappear and the eigenvalues converge. This convergence is indicated by the horizontal alignment of the eigenvalue contours for large $N$. Note that the present achievement of a resolution-independent stability map is distinct from the behavior observed in the case of Perturbation Model I, wherein the required minimum number of Chebyshev modes increased monotically as $N_{min} \approx 2.3\,S^{-1/2}$ (e.g., figure~\ref{fig:couette_Ndependence_combined}). For subsequent analyses, we use $N = 160$ to ensure that the eigenvalues are converged and not numerically spurious.
 
\begin{figure}
    \centering
    \includegraphics[width=0.7\textwidth]{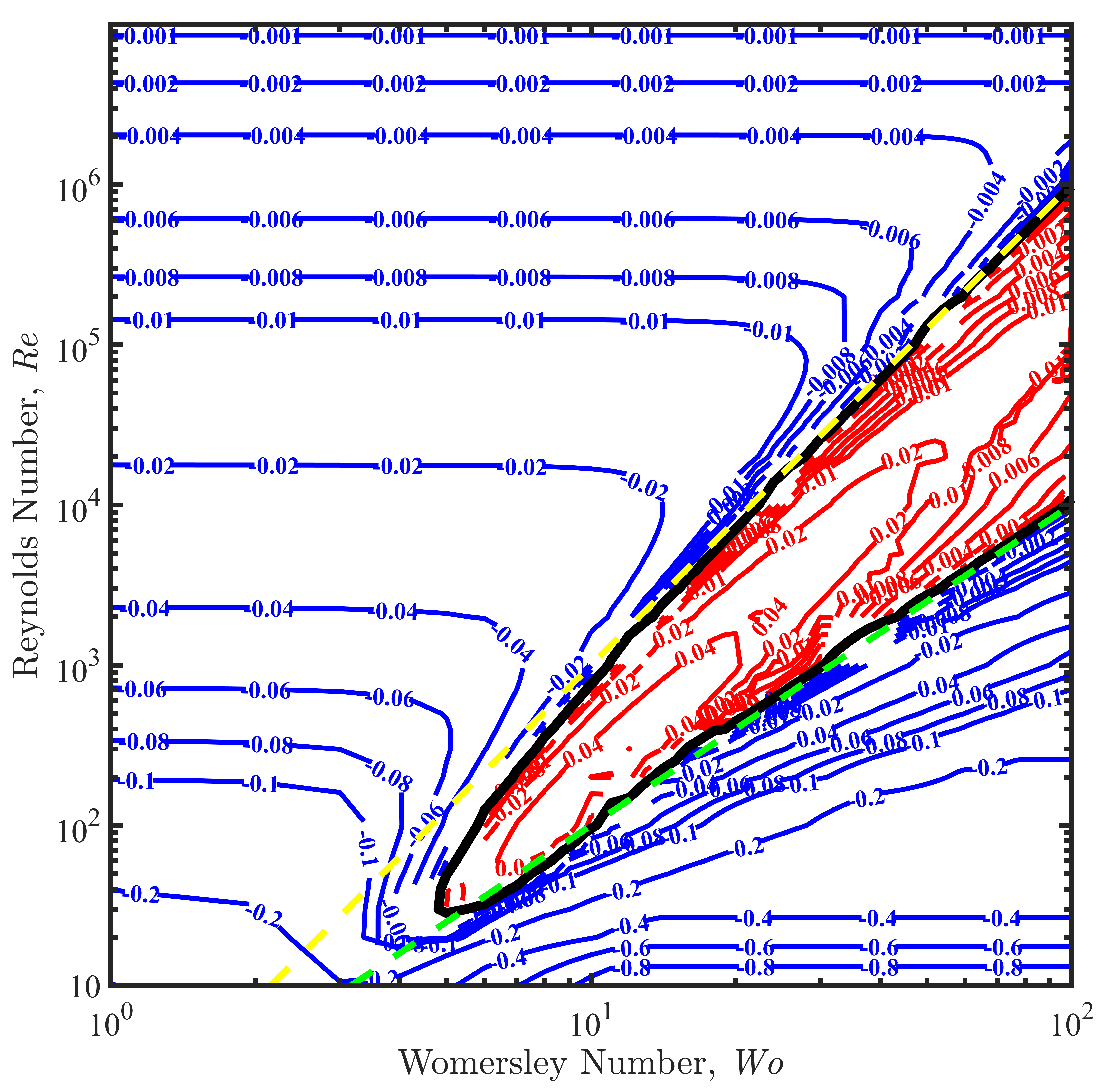}
    \caption{Contour map of the maximum Orr-Sommerfeld eigenvalue imaginary part $\mathbb{I}[\hat{\omega}]$ versus perturbation Womersley number $\widetilde{Wo}$ and Reynolds number $Re$ for plane Couette flow with $(\alpha, \beta) = (1, 0)$. The region of linear instability (red contours) is well described by $\widetilde{Wo}^2 < Re < \widetilde{Wo}^3$, as indicated by the green and yellow dashed lines, respectively.} 
    \label{fig:map_Re_Wo}
\end{figure}

 Intriguingly, a robust band of linear instability is observed in the range $100 < Re < 1000$. To explore this in greater detail, figure \ref{fig:map_Re_Wo} plots a stability map of plane Couette flow versus the perturbation Womersley number $\widetilde{Wo}$ and Reynolds number $Re$ for $(\alpha, \beta) = (1, 0)$. Several interesting features emerge. At Reynolds numbers below $Re \approx 30$, the flow is stable for all values of perturbation Womersley number $\widetilde{Wo}$. Similarly, if the perturbation Womersley number is below $\widetilde{Wo} \approx 5$, the flow is stable for all values of Reynolds number. Both of these effects may be attributable to the dampening effect of the fluid viscosity on the base flow and on the incipient perturbations, respectively. The latter dependence of the flow stability on the nature of the incipient perturbations, as measured by perturbation Womersley number $\widetilde{Wo}$, may explain empirical observations of a range of different Reynolds numbers corresponding to turbulence transition in plane Couette flow experiments \citep{Drazin2004}. Specifically, the present analysis predicts that plane Couette flow becomes linearly unstable for $Re > \widetilde{Wo}^2$, as indicated by the green dashed boundary in figure \ref{fig:map_Re_Wo}.

 Interestingly, this analysis also predicts that linear instability can be avoided at sufficiently high Reynolds numbers if the flow satisfies $Re > \widetilde{Wo}^3$, i.e., the region above the yellow dashed boundary in figure \ref{fig:map_Re_Wo}. This condition may be difficult to achieve in practice for three reasons. First, an arbitrary incipient perturbation may comprise a spectrum of Womersley numbers $\widetilde{Wo}$ that spans a broad range of the abscissa in figure \ref{fig:map_Re_Wo}. Satisfying $Re > \widetilde{Wo}^3$ across that full spectrum may be difficult in the absence of a means to control the frequency spectrum of incipient perturbations \emph{a priori}. Second, a flow generated from rest increases in Reynolds number from zero, and therefore is likely to cross the region of linear instability from below. Third, the condition $Re > \widetilde{Wo}^3$ can be recast in terms of the governing physical parameters as 

\begin{equation}\label{eq:higRestability}
\frac{U_0L_0}{\nu_0} > \frac{\omega^{3/2}}{\nu_0^{3/2}\alpha^3}
\end{equation}

Hence, any attempt to increase the Reynolds number by reducing the viscosity will lead to a concomitant and larger increase in the perturbation Womersley number, making it more difficult to satisfy the stability criterion in equation (\ref{eq:higRestability}).

These challenges notwithstanding, the stability map in figure \ref{fig:map_Re_Wo} is a testable prediction of this model that can be examined in experiments. Confirmation of the model predictions would suggest that this alternative to the no-slip condition may improve the fidelity of linear stability analysis as a predictive tool. Furthermore, if the regime $Re > \widetilde{Wo}^3$ is empirically accessible, it may be possible to achieve stable, laminar flow and associated fluid dynamic performance gains at arbitrarily high Reynolds numbers.

\begin{figure}
    \centering
    \includegraphics[width=0.475\linewidth]{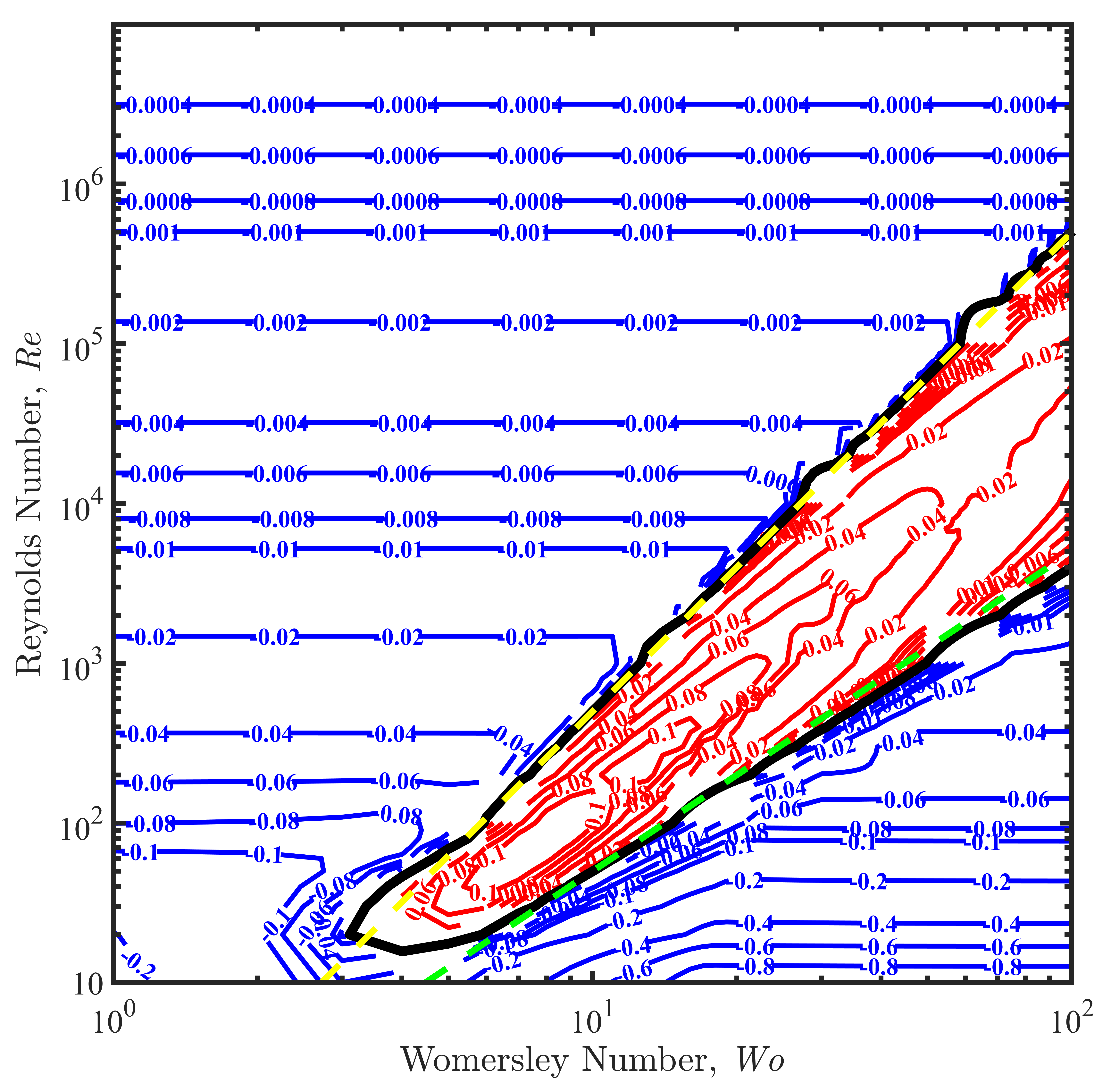}
    \includegraphics[width=0.475\linewidth]{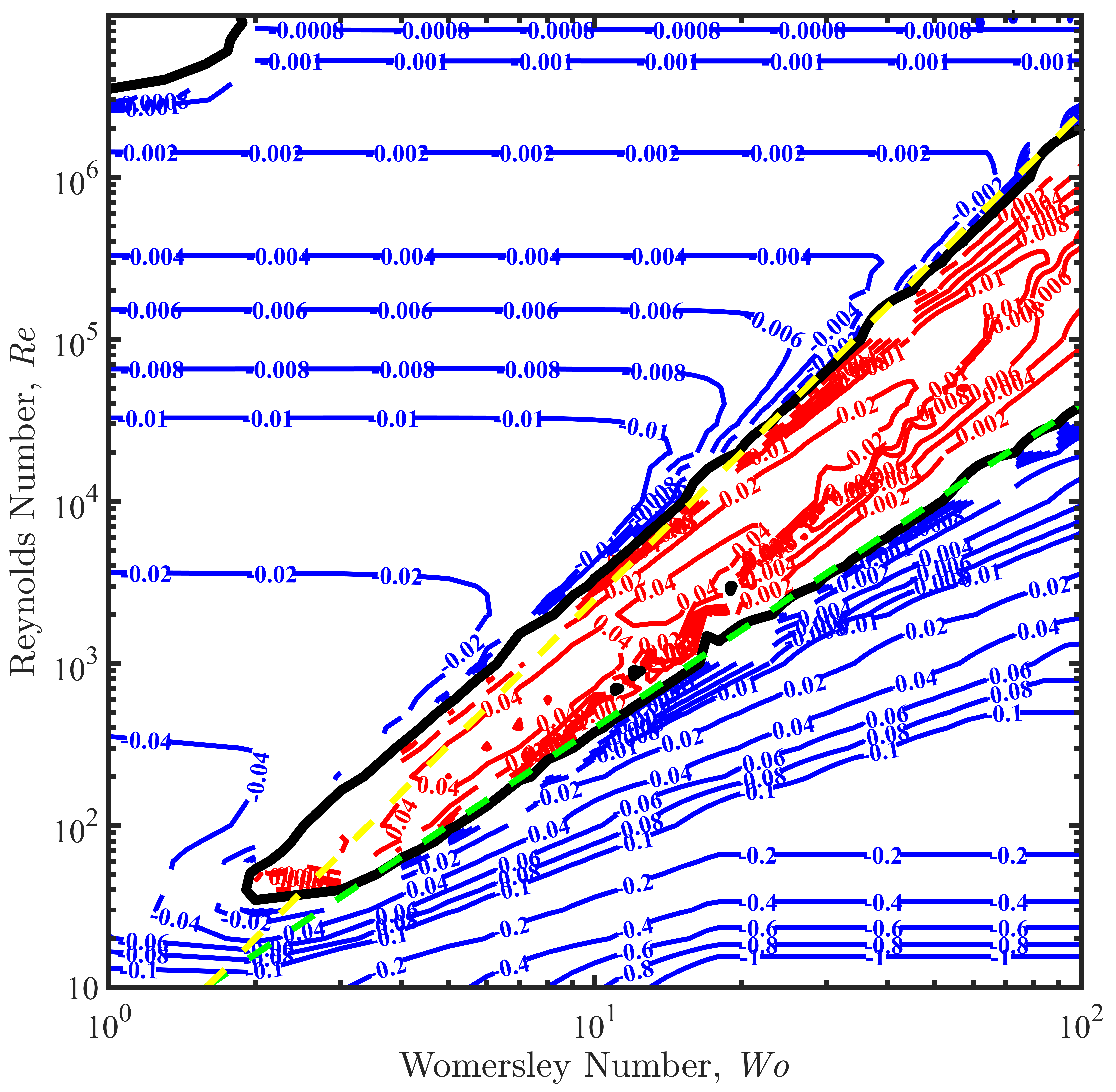}
    \caption{Contour map of the maximum Orr-Sommerfeld eigenvalue imaginary part $\mathbb{I}[\hat{\omega}]$ versus perturbation Womersley number $\widetilde{Wo}$ and Reynolds number $Re$ for plane Poiseuille flow with $(\alpha, \beta) = (1, 0)$ (\emph{left}) and for Hagen-Poiseuille pipe flow with $(\alpha, n) = (1, 1)$ (\emph{right}). The region of linear instability (red contours) is well described by $\widetilde{Wo}^2 < Re < \widetilde{Wo}^3$, as indicated by the green and yellow dashed lines, respectively. See main text for discussion of scaling prefactors.} 
    \label{fig:map_Re_Wo_planepoise}
\end{figure}

Similar results were observed in analysis of plane Poiseuille flow as well as for Hagen-Poiseuille pipe flow (see Appendix A for pipe flow governing equations). As shown in figure \ref{fig:map_Re_Wo_planepoise}, the region of linear instability has a similar shape for both Poiseuille flows as in the plane Couette flow. The neutral stability boundaries have a similar scaling but distinct prefactors, i.e., $0.5\widetilde{Wo}^2 < Re < 0.5\widetilde{Wo}^3$ and $4\widetilde{Wo}^2 < Re < 2.5\widetilde{Wo}^3$, respectively. The  numerical discrepancy in prefactors of the two planar flows can be resolved by defining the Couette flow Reynolds number based on the maximum flow speed contrast in the domain (i.e., $U_0 = 2$) instead of the maximum flow speed; or by using the full-width distance between the walls as the characteristic length instead of the half-width.

A potential conceptual inelegance of the model described herein is its characterization of incipient perturbations using an independent parameter $\widetilde{Wo}$. The frequency $\omega$ used to define the perturbation Womersley number is not necessarily identical to the eigenvalue $\hat{\omega}$ of the Orr-Sommerfeld system. Indeed, the frequency $\omega$ is an input to the boundary condition \ref{eq:womersleybc}, whereas the eigenvalue $\hat{\omega}$ is a solution output. It is unclear whether this distinction is significant, but it is noted here for the sake of completeness.

\section{Conclusions}

The present analyses are motivated by the premise that incipient velocity perturbations in viscous parallel shear flows may not satify the no-slip condition exactly. Two alternative models of the fluid-solid interface physics demonstrated that even small departures from the exact no-slip condition---at levels that may be empirically indistinguishable from the the classical no-slip condition---can significantly affect the predictions of linear stability analysis. Intriguingly, Perturbation Model II predicts a minimum critical Reynolds number for linear instability that is quantitatively consistent with empirical observations of turbulence transition, e.g., transition Reynolds numbers of $\mathcal{O}(10^3-10^4)$ in pipe flow~\citep{Avila2023}. Moreover, the model predicts an unexpected, second regime of high-Reynolds number flow stability. Experimental validation of that prediction would not only provide compelling support for Perturbation Model II as a realistic model of the perturbation flow physics, it would also potentially have important implications in engineering applications.

More generally, the discovered linearly unstable eigenmodes in both perturbation models imply the presence of distinct, near-wall flow kinematics that may also be observable in experiments that confirm their existence. The transient nature of these near-wall fluid particle motions and their appearance as \emph{perturbations} to the base flow may require development of more sensitive measurement techniques than are available currently. Nonetheless, the present theoretical framework can guide experimental efforts to test these concepts further. 

\backsection[Data availability statement]{Matlab implementations of the linear stability codes are openly available in the Caltech Data Repository at https://doi.org/10.22002/4sm2w-br988}

\backsection[Acknowledgements]{The authors thank M. Malik for discussion of the pipe flow calculations and M. K. Fu for manuscript feedback.}

\backsection[Funding]{J.O.D. was supported by the U.S. National Science Foundation.}

\backsection[Declaration of interests]{The authors report no conflict of interest.}

\appendix
\input{appendix1.tex}

\clearpage


\bibliographystyle{jfm}


\end{document}

%% file: appendix1.tex
\section*{Appendix A: Orr-Sommerfeld Equations}\label{appA}
\subsection*{Plane Couette and Plane Poiseuille Flow}

\renewcommand{\theequation}{A.\arabic{equation}}

Following \citet{Schmid2001}, we consider velocity perturbations of the general Cartesian form

\begin{equation}\label{eq:SHperturbationA}
\mathbf{u}(x,y,z,t) = \mathbb{R}[\mathbf{\tilde{u}}(y)e^{i(\alpha x + \beta z - \omega t)}]
\end{equation}

\noindent where the Cartesian components of the velocity vector are $\mathbf{u} = u \mathbf{\hat{i}} + v \mathbf{\hat{j}} + w \mathbf{\hat{k}}$ in the streamwise (i.e., $\mathbf{\hat{i}}$), wall-normal (i.e., $\mathbf{\hat{j}}$), and transverse (i.e., $\mathbf{\hat{k}}$) directions. 
The temporal evolution of the perturbations is examined by treating the spatial wavenumbers as $\alpha,\beta \in \mathbb{R}$ and the frequency as $\omega = \alpha c$, where $c \in \mathbb{C}$ is the complex phase speed of the perturbation. The stability of perturbations of a given spatial wavenumber is evaluated by solving the Orr-Sommerfeld equation:

\begin{equation}\label{eq:OSequation}
\Bigl[(-i\omega + i\alpha U)(\mathcal{D}^2 - k^2)-i\alpha U''-\frac{1}{Re}(\mathcal{D}^2 - k^2)^2\Bigr]\tilde{v} = 0
\end{equation}

\noindent where $k^2 = \alpha^2 + \beta^2$, and $Re = U_0L_0/\nu_0$ is the Reynolds number. Here, $U_0$ is taken as the maximum velocity of the base flow in the domain, and $L_0$ is the half-width of distance between the solid walls of the Couette or Poiseuille flow. The operator $\mathcal{D}$ and the prime both denote a derivative with respect to the wall-normal $y-$coordinate direction. 

For plane Couette flow, the base flow is given by $U(y) = y$ on the domain $-1 \leq y \leq 1$. For plane Poiseuille flow, the base flow is given by $U(y) = 1 - y^2$ on the domain $-1 \leq y \leq 1$. In Perturbation Model I, equation (\ref{eq:OSequation}) is solved using the boundary condition $\tilde{v} = 0$ at $y = \pm1$ (i.e., no flow penetration at the solid walls), along with the generalized boundary condition for the streamwise perturbation component $\tilde{u}$ in equation (\ref{eq:Redependentslip}) to replace the conventional no-slip condition. Given the form of velocity perturbation in equation (\ref{eq:SHperturbationA}), and using $\tilde{u} = (i/\alpha)\mathcal{D}\tilde{v}$ from the continuity equation, the generalized boundary condition for $\tilde{u}$ can be written in terms of the normal velocity $\tilde{v}$ as 

\begin{equation}\label{eq:SHboundarycondition}
\frac{i}{\alpha}\mathcal{D}\tilde{v} \mp \frac{i}{\alpha} S\, \mathcal{D}^2\tilde{v} = 0 \quad \textrm{for} \quad y = \pm1
\end{equation}

In Perturbation Model II, equation (\ref{eq:OSequation}) is solved using boundary conditions (\ref{eq:nothroughflow1}), (\ref{eq:nothroughflow2}), and (\ref{eq:womersleybc}) at $y = \pm1$.

\subsection*{Hagen-Poiseuille Pipe Flow}

Following \citet{Malik2019}, we consider velocity perturbations in cylindrical coordinates of the form

\begin{equation}\label{eq:MSperturbation}
\mathbf{u}(x,r,\theta,t) = \mathbb{R}[\mathbf{\tilde{u}}(r)e^{i(\alpha x + n \theta - \omega t)}]
\end{equation}

\noindent where $(x,r,\theta)$ are the axial, radial, and azimuthal directions, respectively, and $n$ is the azimuthal wavenumber. The Orr-Sommerfeld equations can be expressed as

\begin{equation}\label{eq:MSOSequation}
\begin{split}
& i(\alpha U-\omega)(\mathcal{D}\tilde{u}-i\alpha\tilde{v}) = -i\alpha U'\tilde{u}-(U''+U'\mathcal{D})\tilde{v}+\frac{1}{Re}\Bigl(\Delta(\mathcal{D}\tilde{u}-i\alpha\tilde{v})-2 i n r^{-2}\tilde{\eta}\Bigr) \\
& i(\alpha U-\omega)[\mathcal{D}(r\tilde{w})-in\tilde{v}] = -i\alpha r U'\tilde{w}+\frac{1}{Re}\Bigl(\overline{\Delta}[\mathcal{D}(r\tilde{w})-in\tilde{v}]\Bigr) \\
& i(\alpha U-\omega)\tilde{\eta} = -inU'r^{-1}\tilde{v}+\frac{1}{Re}\Bigl(\Delta\eta + 2nr^{-2}(\alpha\tilde{v}+i\mathcal{D}\tilde{u})\Bigr)
\end{split}
\end{equation}

\noindent where $\Delta = [\mathcal{D}^2+r^{-1}\mathcal{D}-r^{-2}(d+1)]$, $\overline{\Delta} = [\mathcal{D}^2-r^{-1}\mathcal{D}-r^{-2}(d-1)]$, $\tilde{\eta} = i(nr^{-1}\tilde{u}-\alpha\tilde{w})$, $d = n^2 + \alpha^2 r^2$, and the operator $\mathcal{D}$ and the prime both denote a derivative with respect to the radial direction. The base flow for Hagen-Poiseuille pipe flow is given by $U(r) = 1-r^2$ on the domain $0 \leq r \leq 1$.

The Orr-Sommerfeld equations are solved following \citet{Malik2019} by using surrogate analytic functions $(\phi, \Omega)$ that facilitate more straightforward implementation of the perturbation models (see below) and regularity conditions on the axis of symmetry $r = 0$. The wall-normal velocity and vorticity are expressed in terms of these functions as

\begin{equation}\label{eq:MSnormalvelvort}
(\tilde{v}, \tilde{\eta}) = \begin{cases}
			(r^l\phi, r^l\Omega), & n \neq 0\\
            (r\phi, r\Omega), & n = 0
		 \end{cases}
\end{equation}

\noindent where $l = |n| - 1$. The no-penetration boundary condition is therefore given by $\phi = 0$ at $r = 1$.

Using continuity and the normal vorticity definition, the streamwise (i.e., $\tilde{u}$) velocity perturbation component is given by

\begin{equation}\label{eq:MSstreamwisevel}
\tilde{u} = \begin{cases}
			\frac{ir^{l+1}}{d}[\alpha(l+1)\phi + \alpha r \mathcal{D}\phi - n \Omega], & n \neq 0\\
            \frac{i}{\alpha}(2\phi + r\mathcal{D}\phi), & n = 0
		 \end{cases}
\end{equation}

\subsubsection*{Perturbation Model I}

 From equations (\ref{eq:Redependentslip}) and (\ref{eq:MSstreamwisevel}), the generalized boundary condition for the streamwise velocity component at the wall $(r = 1)$ is

\begin{equation}\label{eq:MSReynoldsdependentslip}
\begin{split}
[(a_1 + a_3 + a_6)\mathcal{D} + a_4\mathcal{D}^2]\phi + [(a_2+a_7) + a_5\mathcal{D}]\Omega = 0, \quad n \neq 0\\
[(a_1^0 + a_2^0)\mathcal{D} + a_3^0\mathcal{D}^2]\phi = 0, \quad  n = 0
\end{split}
\end{equation}

\noindent where

\begin{equation}\label{eq:MSReynoldsdependentslipconstants}
\begin{split}
& a_1 = \frac{i\alpha}{d} \\
& a_2 = -\frac{in}{d} \\
& a_3 = -\frac{S\,i\alpha(l+2)}{d} \\
& a_4 = -\frac{S\,i\alpha}{d} \\
& a_5 = \frac{S\,in}{d} \\
& a_6 = -S\,\alpha\Bigl[\frac{i(l+1)d-2i\alpha^2}{d^2}\Bigr] \\
& a_7 = S\,n\Bigl[\frac{i(l+1)d-2i\alpha^2}{d^2}\Bigr]\\
& a_1^0 = \frac{i}{\alpha} \\
& a_2^0 = -3S\frac{i}{\alpha}\\
& a_3^0 = -S\frac{i}{\alpha}\\
\end{split}
\end{equation}

\subsubsection*{Perturbation Model II}

From equations (\ref{eq:stokessecond}) and (\ref{eq:MSstreamwisevel}), the boundary condition for the streamwise velocity component at the wall $(r = 1)$ is

\begin{equation}\label{eq:MSReynoldsdependentslip}
\begin{split}
\Big[a_1\mathcal{D}-\frac{1}{Re}\Big(b_1\alpha+b_2\alpha(l+2)+b_3\alpha(l+2)\Big)\mathcal{D}-\frac{1}{Re}\Big(b_2\alpha+b_3\alpha+b_4\alpha(l+3)\Big)\mathcal{D}^2\\-\frac{1}{Re}\Big(b_4\alpha\Big)\mathcal{D}^3\Big]\phi\\ + \Big[a_2+\frac{1}{Re}\Big(b_1n\Big)+\frac{1}{Re}\Big(b_2+b_3\Big)n\mathcal{D}+\frac{1}{Re}\Big(b_4n\Big)\mathcal{D}^2\Big]\Omega = 0, \quad n \neq 0\\
[(a_1^0 + a_2^0)\mathcal{D} + a_3^0\mathcal{D}^2 + a_4^0\mathcal{D}^3]\phi = 0, \quad  n = 0
\end{split}
\end{equation}

\noindent where

\begin{equation}\label{eq:MSReynoldsdependentslipconstants}
\begin{split}
a_1 = &  \frac{(\alpha \widetilde{Wo})^2}{Re}\Big(\frac{\alpha}{n^2+\alpha^2}\Big)\\
a_2 = &  \frac{-(\alpha \widetilde{Wo})^2}{Re}\Big(\frac{n}{n^2+\alpha^2}\Big)\\
b_1 = & \frac{i(l+1)(2\alpha^2)+i(l+1)^2(n^2+\alpha^2)-2i(l+3)\alpha^2}{(n^2+\alpha^2)^2}\\ - \frac{4\alpha^2\Big(i(l+1)(n^2+\alpha^2)-2i\alpha^2\Big)}{(n^2+\alpha^2)^3} \\
b_2 = & \frac{i(l+1)(n^2+\alpha^2)-2i\alpha^2}{(n^2+\alpha^2)^2}\\
b_3 = & \frac{i(l+2)(n^2+\alpha^2)-2i\alpha^2}{(n^2+\alpha^2)^2}\\
b_4 = & \frac{i}{n^2+\alpha^2}\\
a_1^0 = & \frac{(\alpha\widetilde{Wo})^2}{Re}\\
a_2^0 = & \frac{-3i}{Re}\\
a_3^0 = & \frac{-5i}{Re}\\
a_4^0 = & \frac{-i}{Re}\\
\end{split}
\end{equation}

\section*{Appendix B: Numerical Solution}\label{appB}

The Orr-Sommerfeld eigenvalue equations were solved using Chebyshev collocation as formulated in plane Cartesian coordinates~\citep{Schmid2001} and cylindrical coordinates~\citep{Malik2019}. Matlab implementations of the algorithm for each flow are provided in the Caltech Data Repository.

The Matlab codes were verified by comparing the computed eigenvalues for the case $S = 0$ (i.e., no-slip condition) to the tabulated eigenvalues for plane Couette and Poiseuille flows in \citet{Schmid2001}, as well as the tabulated eigenvalues for Hagen-Poiseuille pipe flow in \citet{Malik2019}.

\subsection*{Perturbation Model I}
For each of the three flows (i.e., plane Couette flow, plane Poiseuille flow, and Hagen-Poiseuille pipe flow), 400 collocation points were used to compute each of the 996 $Re$ $\times$ 66 $S$ $=$ 65,736 total cases in the $S-Re$ parameter space of figure \ref{fig:maps}. Wavenumbers were fixed at $\alpha = 1$ and $\beta = 0$ for plane Couette flow and plane Poiseuille flow, and $\alpha = 1$ and $n = 1$ for Hagen-Poiseuille pipe flow. These wavenumbers typically corresponded to the most unstable eigenmodes, but qualitatively similar results were observed for other wavenumbers. The eigenvalue with maximum imaginary part was recorded for each parameter set $(S, Re)$ studied. If the eigenvalue with second-largest imaginary part was within $10^{-3}$ of the eigenvalue with largest imaginary part, the eigenvalues were treated as a combined pair for the purposes of subsequent analysis.

\subsection*{Perturbation Model II}
For each of the three flows (i.e., plane Couette flow, plane Poiseuille flow, and Hagen-Poiseuille pipe flow), 160 collocation points were used to compute each of the 64 $Re$ $\times$ 100 $\widetilde{Wo}$ $=$ 6,400 total cases in the $\widetilde{Wo}-Re$ parameter space of figures \ref{fig:map_Re_Wo} and \ref{fig:map_Re_Wo_planepoise} . Wavenumbers were fixed at $\alpha = 1$ and $\beta = 0$ for plane Couette flow and plane Poiseuille flow, and $\alpha = 1$ and $n = 1$ for Hagen-Poiseuille pipe flow. These wavenumbers typically corresponded to the most unstable eigenmodes, but qualitatively similar results were observed for other wavenumbers. The eigenvalue with maximum imaginary part was recorded for each parameter set $(\widetilde{Wo}, Re)$ studied. If the eigenvalue with second-largest imaginary part was within $10^{-3}$ of the eigenvalue with largest imaginary part, the eigenvalues were treated as a combined pair for the purposes of subsequent analysis.

\setcounter{equation}{0}
\renewcommand{\theequation}{C.\arabic{equation}}
\setcounter{figure}{0}
\renewcommand{\thefigure}{C.\arabic{figure}}

\section*{Appendix C: Second-Largest Orr-Sommerfeld Eigenvalue Imaginary Part for Perturbation Model I}\label{appC}

\begin{figure*}
    \centering
    \includegraphics[width=\textwidth]{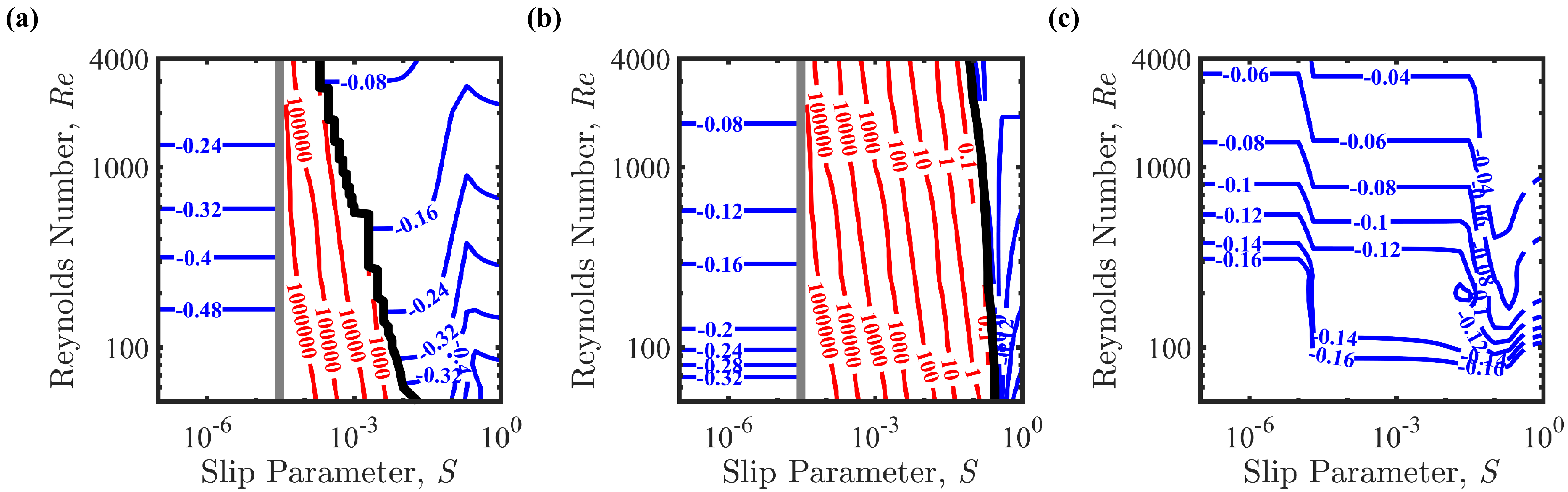}
    \caption{Contour maps of the second-largest Orr-Sommerfeld eigenvalue imaginary part $\mathbb{I}[\hat{\omega}]$ versus slip parameter $S$ and Reynolds number $Re$ for (a) plane Couette flow, (b) plane Poiseuille flow, and (c) Hagen-Poiseuille pipe flow. Blue contours indicate regions of linear stability, and red contours indicate regions of linear instability. Black contours indicate neutral stability boundaries. Vertical gray lines correspond to a discontinuous change in predicted hydrodynamic stability, reflecting the inability of the Chebyshev expansion ($N = 400$) to resolve unstable eigenmodes for values of slip parameter $S$ below the gray line.}
    \label{fig:secondmaps}
\end{figure*}

Figure \ref{fig:secondmaps} plots contours of the second-largest eigenvalue imaginary part for Perturbation Model I. These results demonstrate that the predicted linear instability is not necessarily limited to a single eigenvalue with positive imaginary part.

\setcounter{equation}{0}
\renewcommand{\theequation}{D.\arabic{equation}}

\section*{Appendix D: Asymptotic Analysis of Unstable Couette Eigenmodes for Perturbation Model I}\label{appD}

The Orr-Sommerfeld equation for plane Couette flow can be written as

\begin{equation}\label{eq:OSCouette}
\frac{\partial}{\partial t}\nabla^{2}\tilde{v} - \frac{1}{Re}\nabla^{2}\nabla^{2}\tilde{v}+i\alpha\nabla^{2}\tilde{v} = 0
\end{equation}

\noindent where $\alpha$ is the axial wavenumber. The temporal behavior of linear velocity perturbations is given by $\tilde{v} \sim e^{-i\omega t}$, where the complex frequency $\omega = \mathbb{R}[\omega] + i\mathbb{I}[\omega]$.

We consider the limiting case of large, unstable eigenvalues, i.e., $\mathbb{I}[\hat{\omega}] \gg 1$. To leading order, equation (\ref{eq:OSCouette}) can therefore be written as

\begin{equation}\label{eq:OSCouetteleading}
\Bigg(\mathbb{I}[\hat{\omega}]\,Re + {\alpha}^2 - \frac{\partial^2}{\partial y^2}\Bigg)\nabla^{2}\tilde{v} = 0
\end{equation}

Equation (\ref{eq:OSCouetteleading}) has leading order solutions of the form $\tilde{v} \sim e^{\pm \kappa_1 y},\, e^{\pm \alpha y}$, where $\kappa_1 = \sqrt{\mathbb{I}[\hat{\omega}]\,Re + {\alpha}^2} \gg 1$. The solutions must satisfy a no-penetration condition at the walls, i.e., $\tilde{v}(y = \pm 1) = 0$. This boundary condition is satisfied for

\begin{equation}\label{eq:OSCouetteleadingwithBC1}
\tilde{v}(y) = e^{\kappa_1 (y-1)} - \frac{\sinh \alpha(y+1)}{\sinh 2\alpha} + \kappa_2 \Bigg(e^{-\kappa_1(y+1)} + \frac{\sinh \alpha(y-1)}{\sinh 2\alpha}\Bigg)
\end{equation}

\noindent since $\tilde{v}(y = \pm 1) \sim \mathcal{O}(e^{-2\kappa_1}) \approx 0$. The free parameter $\kappa_2$ can be constrained by applying the boundary condition in equation (\ref{eq:Redependentslip}) as formulated for plane Couette flow in equation (\ref{eq:SHboundarycondition}). Specifically, at $y = 1$:

\begin{equation}\label{eq:OSCouetteleadingwithBC2a}
\frac{\kappa_1 - \alpha \frac{\cosh 2\alpha}{\sinh 2\alpha} + \frac{\alpha \kappa_2}{\sinh 2\alpha}}{\kappa_1^2-\alpha^2 } = S
\end{equation}

\noindent where terms of $\mathcal{O}(e^{-2\kappa_1})$ or smaller have been neglected. Similarly, at $y = -1$:

\begin{equation}\label{eq:OSCouetteleadingwithBC2b}
\frac{-\frac{\alpha}{\sinh \alpha} - \kappa_1 \kappa_2 + \alpha \kappa_2 \frac{\cosh 2\alpha}{\sinh 2\alpha}}{\kappa_2 (\kappa_1^2-\alpha^2)}  = -S
\end{equation}

Combining equations (\ref{eq:OSCouetteleadingwithBC2a}) and (\ref{eq:OSCouetteleadingwithBC2b}) gives $\kappa_2 = \pm 1$. The negative solution $\kappa_2 = -1$ gives antisymmetric eigenmodes consistent with the opposite direction of travel of the walls at $y = \pm 1$. Substituting this solution for $\kappa_2$ into the system of equations gives the approximation $\kappa_1 \approx S^{-1}$, or equivalently:

\begin{equation}\label{eq:omegaeqn}
\mathbb{I}[\hat{\omega}] \sim S^{-2}\,Re^{-1}
\end{equation}

This predicted scaling is consistent with the unstable eigenvalues computed using Chebyshev collocation with $N = 400$ in figure \ref{fig:couette_model_combined}(a) for $Re = 50$, 360, and 3000. The preceding asymptotic analysis also predicts that eigenmode profiles corresponding to large, unstable eigenvalues have the approximate shape given in equations (\ref{eq:unstablemode_u}) and (\ref{eq:unstablemode_v}). This prediction is in excellent agreement with the profiles computed using Chebyshev collocation, as shown in figure \ref{fig:couette_model_combined}(b).

\setcounter{equation}{0}
\renewcommand{\theequation}{E.\arabic{equation}}
\setcounter{figure}{0}
\renewcommand{\thefigure}{E.\arabic{figure}}

\section*{Appendix E: Alternative Boundary Conditions for Perturbation Model I}

\begin{figure*}
    \centering
    \includegraphics[width=\textwidth]{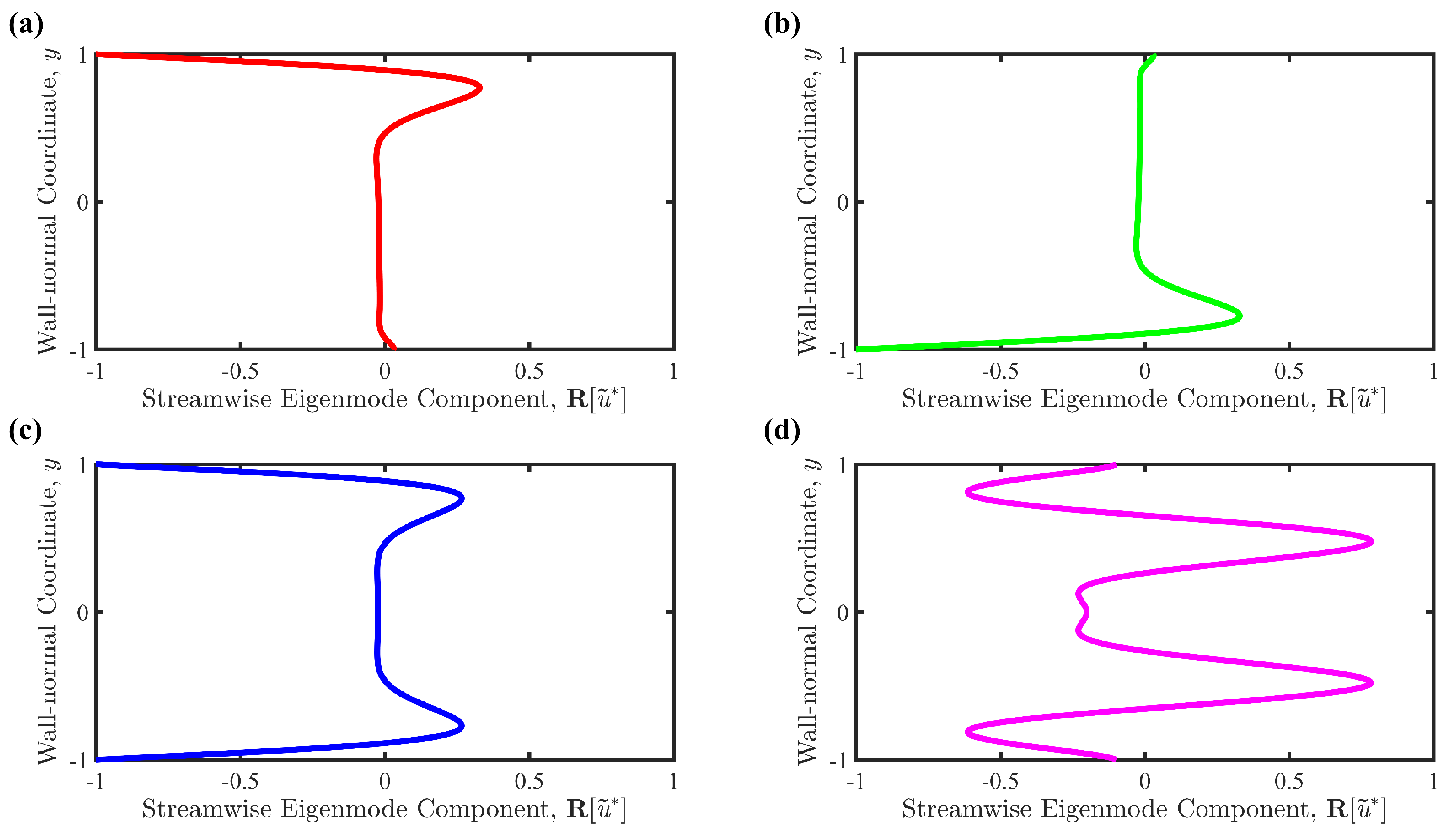}
    \caption{Streamwise (i.e., $\mathbb{R}[\tilde{u}]$) eigenmode components for eigenvalues with maximum imaginary part in plane Couette flow ($Re = 360$, $S = 1 \times 10^{-1}$) with generalized boundary condition in equation (\ref{eq:Redependentslip}) (i.e., panel (c)) and alternative boundary conditions in equation (\ref{eq:alternativeBCs}) (i.e., panels (a,b,d)). Directional alignment of perturbation wall slip and associated wall shear is enforced on (a) top wall; (b) bottom wall; (c) both walls; (d) neither wall. Eigenmodes in panels (a-c) are unstable; eigenmode in panel (d) is stable.}  
    \label{fig:SI_modes_couette_ucomp}
\end{figure*}

To explore the dependence of the discovered unstable eigenmodes in Perturbation Model I on the form of the boundary condition ansatz in equation (\ref{eq:Redependentslip}), Figure \ref{fig:SI_modes_couette_ucomp} plots the streamwise component of the eigenmode with maximum imaginary part for $Re = 360$ and $S = 1 \times 10^{-1}$, using the following alternative boundary conditions:

\begin{equation}\label{eq:alternativeBCs}
\begin{split}
\text{Panel (a):} \quad \tilde{u}(\mathbf{x_{wall}},t) - S\, \tilde{u}'(\mathbf{x_{wall}},t) = 0, & \quad y = \pm 1\\
\text{Panel (b):} \quad \tilde{u}(\mathbf{x_{wall}},t) + S\, \tilde{u}'(\mathbf{x_{wall}},t) = 0, & \quad y = \pm 1\\
\text{Panel (c):} \quad \tilde{u}(\mathbf{x_{wall}},t) \mp S\, \tilde{u}'(\mathbf{x_{wall}},t) = 0, & \quad y = \pm 1\\
\text{Panel (d):} \quad \tilde{u}(\mathbf{x_{wall}},t) \pm S\, \tilde{u}'(\mathbf{x_{wall}},t) = 0, &  \quad y = \pm 1\\
\end{split}
\end{equation}

The boundary condition in panel (a) constrains the streamwise component $\tilde{u}$ of any velocity perturbation at the top wall ($y = 1$) to be oriented in the same direction as the associated shear exerted by the wall on the fluid due to the perturbation. Conversely, on the bottom wall ($y = -1$), the boundary condition sets the direction of any velocity perturbation to be opposite to the direction of the associated shear exerted by the wall on the fluid. The boundary condition in panel (b) is the inverse of panel (a), constraining the streamwise component $\tilde{u}$ of any velocity perturbation at the bottom wall ($y = -1$) to be oriented in the same direction as the associated shear exerted by the wall on the fluid due to the perturbation. The boundary condition on the top wall in panel (b) is identical to the boundary condition on the bottom wall in panel (a). Panel (c) illustrates a boundary condition that is symmetric with respect to its treatment of the top and bottom walls, as in equation (\ref{eq:Redependentslip}). In this case the streamwise component of velocity perturbations at both walls is constrained to be oriented in the same direction as the associated shear exerted by the wall on the fluid due to the perturbation. In panel (d), the boundary condition is also symmetric with respect to its treatment of the top and bottom walls. However, the streamwise component of velocity perturbations at both walls is constrained to be oriented in the direction opposite to the associated wall shear. This boundary condition is similar to that used in prior studies of the effect of wall slip on linear instability (e.g.,~\cite{Lauga2005, Chai2019, Ceccacci2022}). 

\begin{figure*}
    \centering
    \includegraphics[width=\textwidth]{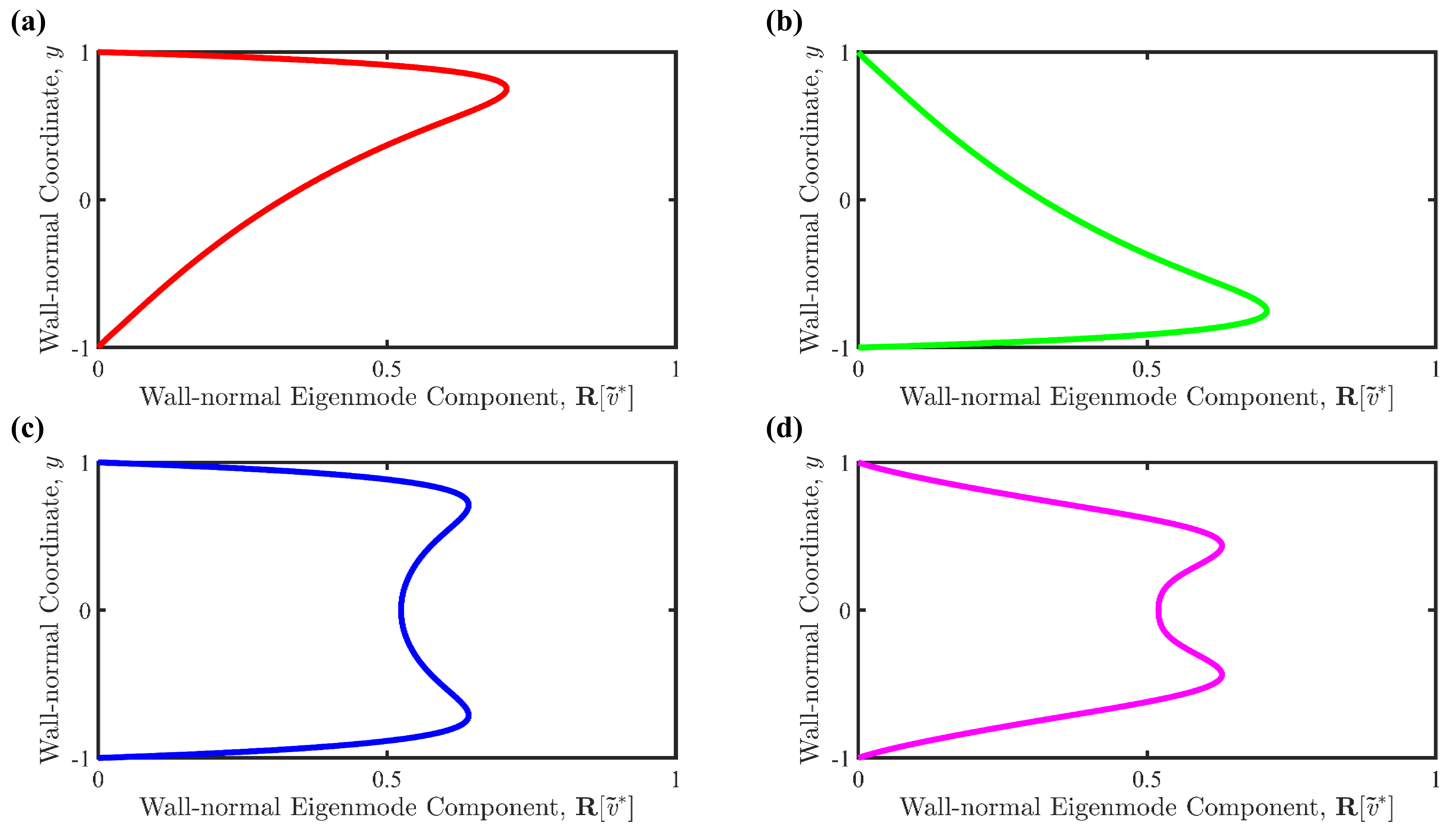}
    \caption{Wall-normal (i.e., $\mathbb{R}[\tilde{v}]$) eigenmode components for eigenvalues with maximum imaginary part in plane Couette flow ($Re = 360$, $S = 1 \times 10^{-1}$) with generalized boundary condition in equation (\ref{eq:Redependentslip}) (i.e., panel (c)) and alternative boundary conditions in equation (\ref{eq:alternativeBCs}) (i.e., panels (a,b,d)). Large near-wall $\tilde{v}$-component perturbations arise near walls with boundary condition enforcing directional alignment of perturbation wall slip and associated wall shear. Specifically, directional alignment of perturbation wall slip and associated wall shear is enforced on (a) top wall; (b) bottom wall; (c) both walls; (d) neither wall. Eigenmodes in panels (a-c) are unstable; eigenmode in panel (d) is stable.}  
    \label{fig:SI_modes_couette_vcomp}
\end{figure*}

Figure \ref{fig:SI_modes_couette_vcomp} shows that the corresponding wall-normal component of velocity perturbations, while satisfying the no-penetration condition at the walls (i.e., $\tilde{v} = 0$), is biased toward walls for which the wall slip of the perturbation is aligned with the wall shear exerted on the fluid due to the velocity perturbations. 

Notably, each boundary condition for which the velocity perturbation and wall shear are aligned on at least one wall (i.e., panels (a), (b), and (c)) corresponds to an unstable eigenmode. In each of the present cases, the growth rate of the perturbation is $\mathbb{I}[\hat{\omega}] \approx 0.199$. By contrast, for the boundary condition in panel (d), all eigenmodes are stable, in agreement with previous studies (e.g.,~\cite{Lauga2005, Chai2019, Ceccacci2022}). The least stable eigenmode has a growth rate $\mathbb{I}[\hat{\omega}] \approx -0.175$.

